\documentclass[a4paper,twoside]{article}
\usepackage[authoryear]{natbib}
\bibpunct{(}{)}{;}{a}{}{,}
\usepackage{graphicx}
\date{} %Please leave the date blank
\title{\large\bf\flushleft Triaxial vs. Spherical Dark Matter Halo Profiles}
\author{\parbox{\textwidth}{\flushleft
\vspace{-0.5cm}
{\it Alexander Knebe and Volkmar Wie\ss ner}\\
\vspace{0.4cm}
{\small Astrophysikalisches Institut Potsdam, An der Sternwarte 16, 14482 Potsdam, Germany}\\
{\small Email: aknebe@aip.de}}}
\begin{document}
\twocolumn[
\begin{changemargin}{.8cm}{.5cm}
\begin{minipage}{.9\textwidth}
\vspace{-1cm}
\maketitle
%
%
%%%%%%%%%%%%%     ABSTRACT    %%%%%%%%%%%%%
%Abstract of no more than 200 words here.
\small{\bf Abstract:}
When analysing dark matter halos forming in cosmological $n$-body simulations it is common practice to obtain the density profile utilizing spherical shells. However, it is also known that the systems under investigation are far from spherical symmetry and rather follow a triaxial mass distribution. In this study we present an estimator for the error introduced by spherically averaging an elliptical mass distribution. We systematically investigate the differences arising when using a triaxial density profile under the assumption of spherical symmetry. We show that the variance in the density can be as large as 50\% in the outer parts of dark matter halos for extreme (but still credible) axis ratios of $0.55:0.67:1$. The inner parts are less affected but still show a scatter at the 16\% level for these prolate systems. For more moderate ellipticities, i.e. axis ratios of $0.73:0.87:1$, the error is smaller but still as large as 10-20\% depending on distance. We further provide a simple formula that allows to estimate this variance as a function of radius for arbitrary axis ratios. We conclude that highly prolate and/or oblate systems are better fit by analytical profiles that take into account the triaxial nature of cosmological objects.

%%%%%%%%%%%%%     KEYWORDS    %%%%%%%%%%%%%
\medskip{\bf Keywords:} cosmology: theory --- cosmology: dark matter --- methods: analytical --- methods: $n$-body simulations
% Please write all keywords in lower case. PASA uses the
% standard list of subject headings adopted by The Astrophysical Journal
% and available from http://www.journals.uchicago.edu/ApJ/keywords_text.html.
% Keywords are separated by em-dashes, i.e. ---

%%%%%%%%DO NOT EDIT%%%%%%%%%%%%
\medskip
\medskip
\end{minipage}
\end{changemargin}
]
\small
%%%%%%%%EDIT FROM HERE%%%%%%%%%%%%
%Please see the PASA Style Guide for help with correct layout for your manuscript.
%Examples of tables and figures are given below.

%%%%%%%%%%%%%%%%%%%%%%%%%%%%%%%%%%%%%%%%%%%%%%%%%%%%%%%%%%%%%%%%%%%%%%%%%%%%%%%%%%%%%%%
\section{Introduction}
%%%%%%%%%%%%%%%%%%%%%%%%%%%%%%%%%%%%%%%%%%%%%%%%%%%%%%%%%%%%%%%%%%%%%%%%%%%%%%%%%%%%%%%

The current standard model of structure formation, in which the Universe is dominated by cold dark matter, predicts that the radial density profile of dark matter halos is best described by the functional form proposed by \citet{Navarro1997} (hereafter NFW):

\begin{equation} \label{NFWspherical}
 \rho(r) = \frac{\rho_c}{\left( r/r_s \right)^\alpha(1+r/r_s)^{3-\alpha}} \ ,
\end{equation}

\noindent
with the two fitting parameters $\rho_c$ and $r_s$, and $\alpha$ was found to be equal to unity, i.e. $\alpha=1$.

During the last decade great strides have been undertaken to verify and better understand the origin of this density profile of dark matter halos forming in cosmological $n$-body simulations. The functional form proposed by NFW is independent of cosmology and the two fitting parameters $\rho_c$ and $r_s$ are actually related.  Equation~\ref{NFWspherical} therefore describes a one-parameter family of functions. While \cite{Navarro1997} still relied on systems containing of the order $10^4$ particles (and hence resolving the profile down to about 10\% of the virial radius, according to the convergence criteria of \citet{Power2003}) studies containing about two orders of magnitude more particles within the virial radius were only able to confirm their findings (e.g. \citealt*{Fukushige1997}; \citealt*{Fukushige2001}; \citealt*{Moore1998}; \citealt*{Jing2000}; \citealt*{JingSuto2000}; \citealt*{Jing2002}). However, there has been a debate about the exact value of the logarithmic inner slope ranging from $\alpha \sim 1$~to~$1.5$ (e.g., \citealt{Navarro1997}, \citealt{Moore1999}). Recently a great deal of effort has gone into even higher resolution simulations which have revealed density profiles of cold dark matter halos down to scales well below one percent of the virial radius (\citealt*{Fukushige2004}; \citealt*{Tasitsiomi2004}; \citealt*{Navarro2004}; \citealt*{Reed2005}; \citealt*{Diemand2004}). The highest resolved simulation to-date reached  an effective mass resolution of about 130 million particles in a cluster sized dark matter halo still supporting the evidence for a central cusp with a logarithmic inner slope of about $\alpha \sim 1.2$ (\citealt*{Diemand2005}). However, these steep central density cusps are not supported by observations. By measuring the rotation curve of a galaxy, one can in principle infer the density profile of its dark matter halo. High resolution observations of low surface brightness galaxies are though best fit by halos with a core of constant density (Simon~et al. 2005; Swaters~et al. 2003; de Block \& Bosma 2002). This discrepancy between the prediction of theory and actual observations has caught the attention of cosmologists for nearly a decade now and is far from being resolved.

Aside from addressing the mass distribution of dark matter halos from a simulator's point of view, a fair amount of labour went into a theoretical understanding of the origin of the NFW profile equation~(\ref{NFWspherical}). But only when making strong assumptions about a connection between the density and the velocity dispersion can one can make analytical predictions. For example, under the assumption that the phase-space density is a power law in radius as suggested by \cite{Taylor2001} one can solve the spherical Jeans equation to obtain the density profile (\citealt*{Dehnen2005}, \citealt*{Austin2005}, \citealt*{Hansen2004}). These studies, guided by the results of the numerical simulations, confirm the inner logarithmic slope to be in the range $\alpha \sim 1$ to $2$. But see also \cite{Hansen2006} for recent claims that equilibrated dark matter structures have a central slope of $\alpha \sim 0.8$.

It needs to mentioned that (nearly) all these studies are based upon the assumption of spherical symmetry utilizing radially distributed shells. But \cite{Jing2002} already noted that halo density profiles as found in cosmological simulations are better fit by applying an elliptical or triaxial modeling of dark matter halos. Detailed investigations of the distributions of shapes of halos forming in cosmological $n$-body simulations now demonstrate that the majority of objects has intrinsic shapes best characterized by axis ratios of about 1:0.74:0.64 (e.g., \citealt{Kasun2005}; \citealt{Bailin2005}, \citealt{Allgood2006}). \cite{Oguri2005}, for instance, showed that such prolate mass distributions can have a significant influence on the interpretation of lensing signals observed in galaxy clusters, i.e. halo triaxiality may act to bias the profile constraints derived under the assumption of a spherically symmetric mass distribution. It therefore appears mandatory to systematically study the error introduced by the supposition that halo profiles are spherical while their actual distribution is triaxial. Can the debate about the central cusp be connected to this error? Answering this question is the major motiviation for the current study.

%%%%%%%%%%%%%%%%%%%%%%%%%%%%%%%%%%%%%%%%%%%%%%%%%%%%%%%%%%%%%%%%%%%%%%%%%%%%%%%%%%%%%%%
\section{The Equations}
%%%%%%%%%%%%%%%%%%%%%%%%%%%%%%%%%%%%%%%%%%%%%%%%%%%%%%%%%%%%%%%%%%%%%%%%%%%%%%%%%%%%%%%
A model for a triaxial NFW density profile is given by, for instance, \cite{Jing2002} and reads as follows:

\begin{equation} \label{NFWtriaxial}
 \varrho(R) = \frac{\varrho_c}{R/R_s (1+R/R_s)^2} \ ,
\end{equation}

\noindent
where the relation between the spherical and elliptical coordinates for a given Cartesian point $x,y,z$

\begin{equation} \label{SEtrafo}
 \begin{array}{rclcl}
  x & = & r \sin{\theta} \cos{\phi} & = & R \frac{a}{c} \sin{\Theta} \cos{\Phi}\\
  y & = & r \sin{\theta} \sin{\phi} & = & R \frac{b}{c} \sin{\Theta} \sin{\Phi}\\
  z & = & r \cos{\theta}            & = & R             \cos{\Theta}   \\  
 \end{array}
\end{equation}

\noindent
depends on the axes $a<b<c$ of the ellipsoid and hence the degree of triaxiality~$T$ (e.g. \citealt{Franx1991})

\begin{equation}
 T = \frac{c^2-b^2}{c^2-a^2} \ .
\end{equation}

If we now average the triaxial density $\varrho(R)$ on the surface of a sphere with radius $r$

\begin{equation} \label{meanRho}
 \langle\varrho(r)\rangle_{r={\rm const}} = \frac{1}{4\pi} \int_{\theta=0}^{\pi} \int_{\phi=0}^{2\pi} \varrho(R) \sin{\theta} d\theta d\phi \ ,
\end{equation}

\noindent
where $\varrho(R)=\varrho(R(r,\theta,\phi))$ is a function of $r, \theta, {\rm and\ } \phi$ through equation~(\ref{SEtrafo}), we can obtain a measure for the error for the variation of the actual density $\varrho(R)$ by calculating the dispersion

\begin{equation} \label{error}
 \sigma^2(r) = \frac{1}{4\pi}  \int_{\theta=0}^{\pi} \int_{\phi=0}^{2\pi}
               \left|\frac{\varrho(R)-\langle\varrho(r)\rangle}{\langle\varrho(r)\rangle}\right|^2 \sin{\theta} d\theta d\phi \ .
\end{equation}

We need to stress that this quantity is per definition the typical dispersion in the density for points on the surface of a sphere of a given radius due to the ellipticity of the halo. There are hence parts on the sphere with values larger than the spherically averaged density as well as parts with lower densities. However, we though refer to $\sigma^2(r)$ as the "error" measure (or error estimate) as it quantifies the mean deviation (either positive or negative) from the actual (triaxial) density distribution under the assumption of spherical symmetry.

\begin{figure}[h]
\begin{center}
\includegraphics[scale=0.35, angle=0]{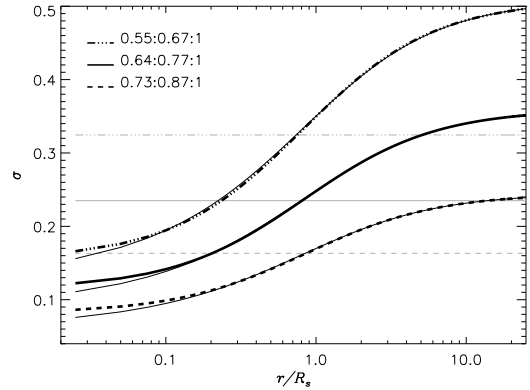}
\caption{The error estimate equation~(\ref{error}) (thick lines) for the most likely axis ratio $a:b:c$ (solid line) and the $1-\sigma$ scatter (dot-dashed and dashed lines) as derived from cosmological $n$-body simulations. The faint horizontal lines correspond to the error when using an isothermal density profile as given by equation~(\ref{isothermal}).}\label{fig:error}
\end{center}
\end{figure}

\begin{figure}[h]
\begin{center}
\includegraphics[scale=0.35, angle=0]{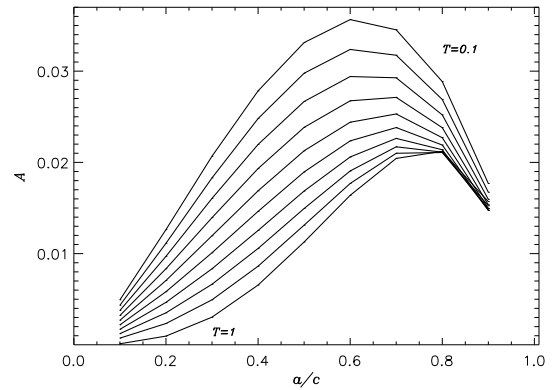}
\caption{Variation of fitting parameter $A$ with $a/c$ along lines of constant triaxiality $T$.}\label{fig:fitAmp}
\end{center}
\end{figure}

\begin{figure}[h]
\begin{center}
\includegraphics[scale=0.35, angle=0]{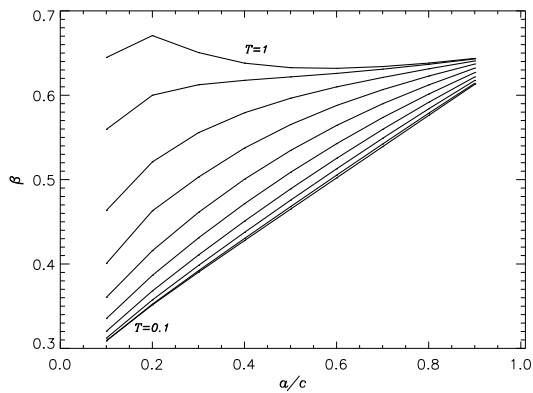}
\caption{Same as figure~\ref{fig:fitAmp} but showing $\beta$ as a function of $a/c$.}\label{fig:fitbeta}
\end{center}
\end{figure}

%%%%%%%%%%%%%%%%%%%%%%%%%%%%%%%%%%%%%%%%%%%%%%%%%%%%%%%%%%%%%%%%%%%%%%%%%%%%%%%%%%%%%%%
\section{The Results}
%%%%%%%%%%%%%%%%%%%%%%%%%%%%%%%%%%%%%%%%%%%%%%%%%%%%%%%%%%%%%%%%%%%%%%%%%%%%%%%%%%%%%%%
In figure~\ref{fig:error} we now plot the error estimate as given by equation~(\ref{error}) (thick lines) for a number of axis ratios. The ratios chosen agree with the findings of cosmological $n$-body simulations as reported by \cite{Kasun2005}. Their distributions are close to Gaussian and we chose to present equation~(\ref{error}) for axis ratios based upon the mean ratios $\langle a/c \rangle$ and $\langle b/c \rangle$ as well as the $1\sigma$ scatter about these values. This leaves us with the ratios $\langle a/c \rangle = 0.64 \pm 0.09$ and $\langle b/c \rangle = 0.77 \pm 0.10$. Hence we plot equation~(\ref{error}) for the ratios 

\begin{itemize}
 \item $a:b:c=0.64:0.77:1$ (mean),
 \item $a:b:c=0.73:0.87:1$ ($+1\sigma$), and
 \item $a:b:c=0.55:0.67:1$ ($-1\sigma$).
\end{itemize}

While the errors are rather small in the central parts we note a discrepancy of up to 50\% in the outer regions of dark matter halos for the more triaxial axis ratios.

To better understand and gauge the functional behaviour of the curves presented in figure~\ref{fig:error} we calculated the error estimate equation~(\ref{error}) for the same axis ratios but using a simple isothermal density profile

\begin{equation} \label{isothermal}
 \varrho(R) = \frac{\varrho_c}{(R/R_s)^2} \ .
\end{equation}

The results can be viewed in figure~\ref{fig:error} too, and are represented by faint horizontal lines for the respective axis ratios. We note that for a single power-law density profile the error is constant with radius. We further observe that the crossing point between the isothermal and the NFW profiles occurs about the scale radius (i.e. $r/R_s \sim 1$) at which the NFW profile approximates the isothermal behaviour. This suggests to fit the radial dependence of the NFW error estimate by a curve with a functional form similar to the logarithmic slope of the density profile itself, i.e. ${d\log\varrho}/{d\log R}$. We found the following function to accurately describe the data

\begin{equation} \label{sigfit}
 \sigma_{\rm fit}(r) = A \left[\frac{c^2}{ab}\left( 1+\left(\frac{r/R_s}{1+r/R_s}\right)^\beta  \right)\right]^2 \ ,
\end{equation}

\noindent
where $A$ and $\beta$ are free parameters. The best fit curves to the error equation~(\ref{error}) are also shown in figure~\ref{fig:error} as thin solid lines.

The question now rises about the dependency of the fitting parameters $A,\beta$ on the axis ratios $a:b:c$ and the triaxiality $T$, respectively. To this extent we show in figures~\ref{fig:fitAmp} and \ref{fig:fitbeta} how $A$ and $\beta$ vary as a function of $a/c$ along lines of constant triaxiality $T$. We note that the variation in both parameters is rather small and the average value of the median values for constant $T$ amount to $\langle A\rangle=0.017$ and $\langle \beta\rangle=0.530$. When adopting these parameters, equation~(\ref{sigfit}) provides a (universal) formula to estimate the error introduced by spherically averaging an elliptical mass distribution for given axes ratios $a/c$ and $b/c$, respectively.

%%%%%%%%%%%%%%%%%%%%%%%%%%%%%%%%%%%%%%%%%%%%%%%%%%%%%%%%%%%%%%%%%%%%%%%%%%%%%%%%%%%%%%%
\section{Conclusions}
%%%%%%%%%%%%%%%%%%%%%%%%%%%%%%%%%%%%%%%%%%%%%%%%%%%%%%%%%%%%%%%%%%%%%%%%%%%%%%%%%%%%%%%
Motivated by the findings of cosmological simulations that dark matter halos have triaxial mass distributions (e.g., \citealt{Kasun2005}, \citealt{Bailin2005}, \citealt{Allgood2006}) we investigated the errors in the derivation of the density profile under the assumption of spherical symmetry. This supposition is rather common and nearly all results presenting evidence for the "cuspy" behaviour of dark matter halos rely upon this premise. While others already showed that a spherical assumption will lead to discrepancies in the interpretation of observational data (e.g., \citealt{Jing2002}; \citealt{Oguri2005}) we presented an estimator to quantify this error. 

We set out to shed light on the question whether the discrepancy and debate about the exact value of the logarithmic central slope (e.g., \citealt{Navarro1997}, \citealt{Moore1999}) can be related to the error introduced by the simplifying assumption of spherical symmetry. We now conclude that the inner parts are only marginally affected and the error will be too small to explain variations in the slope as large as $\pm 0.5$. A change in $\alpha$ from, for instance, $1.5$ down to $1$ entails a difference in the central density of a factor of about 3 (at approx. 10\% of the scale radius) rather than the low percentage level indicated by equation~(\ref{error}) and seen in figure~\ref{fig:error}. However, there are claims that dark matter halos tend to be more elliptical in the central regions (\citealt{Dubinski1991}, \citealt{Jing2002}, \citealt{Kazantzidis2004}, \citealt{Bailin2005b}, \citealt{Allgood2006}, \citealt{Power2006}). But those shape changes are only minor and will hence not be able to lift the error to the required level.

We further provided a fitting function that allows to estimate the error introduced by spherically averaging a triaxial mass distribution as a function of radius for (almost) arbitrary axis ratios $a/c$ and $b/c$.

%%Format tables as in the following example
%\begin{table}[h]
%\begin{center}
%\caption{Example Table}\label{tableexample}
%\begin{tabular}{lcc}
%\hline Column 1 & Column 2 & Column 3 \\
%\hline Table Content$^a$ \\
%\hline
%\end{tabular}
%\medskip\\
%$^a$Table footnotes go here.\\
%\end{center}
%\end{table}

%%%%%%%%%%%%%%%%%%%%%%%%%%%%%%%%%%%%%%%%%%%%%%%%%%%%%%%%%%%%%%%%%%%%%%%%%%%%%%%%%%%%%%%
\section*{Acknowledgments} %If needed
%%%%%%%%%%%%%%%%%%%%%%%%%%%%%%%%%%%%%%%%%%%%%%%%%%%%%%%%%%%%%%%%%%%%%%%%%%%%%%%%%%%%%%%
We benefitted from valuable discussions with Volker M\"uller. AK acknowledges funding through the Emmy Noether Programme by the DFG (KN 755/1).

%%%%%%%%%%%%%%%%%%%%%%%%%%%%%%%%%%%%%%%%%%%%%%%%%%%%%%%%%%%%%%%%%%%%%%%%%%%%%%%%%%%%%%%

%\end{multicols}

\end{document}